\documentclass[
 reprint,
 amsmath,amssymb,
 aps,
pra,
]{revtex4-2}
\usepackage{graphicx}
\usepackage{float}
\usepackage{dcolumn}
\usepackage{bm}
\usepackage{bbm}
\usepackage{braket}
\usepackage{xcolor}
\usepackage[T1]{fontenc}
\usepackage{enumerate,enumitem}
\usepackage{booktabs}
\usepackage{physics}

\begin{document}

\title{Flying-cat parity checks for quantum error correction}

\author{Z.~M.~McIntyre }
\email{zoe.mcintyre@mail.mcgill.ca}
\author{W.~A.~Coish}%
 \email{william.coish@mcgill.ca}
\affiliation{%
 Department of Physics, McGill University, 3600 rue University, Montreal, QC, H3A 2T8, Canada
}%

\date{\today}

\begin{abstract}
   Long range, multi-qubit parity checks have applications in both quantum error correction and measurement-based entanglement generation.  Such parity checks could be performed using  qubit-state-dependent phase shifts on propagating pulses of light described by coherent states $\ket{\alpha}$ of the electromagnetic field. We consider ``flying-cat'' parity checks based on an entangling operation that is quantum non-demolition (QND) for Schr\"odinger's cat states $\ket{\alpha}\pm \ket{-\alpha}$. This operation encodes parity information in the phase of maximally distinguishable coherent states $\ket{\pm \alpha}$, which can be read out using a phase-sensitive measurement of the electromagnetic field. In contrast to many implementations, where single-qubit errors and measurement errors can be treated as independent, photon loss during flying-cat parity checks introduces errors on physical qubits at a rate that is anti-correlated with the probability for measurement errors. We analyze this trade-off for three-qubit parity checks, which are a requirement for universal fault-tolerant quantum computing with the subsystem surface code. We further show how a six-qubit entangled ``tetrahedron'' state can be prepared using these three-qubit parity checks. The tetrahedron state can be used as a resource for controlled quantum teleportation of a two-qubit state, or as a source of shared randomness with potential applications in three-party quantum key distribution. Finally, we provide conditions for performing high-quality flying-cat parity checks in a state-of-the-art circuit QED architecture, accounting for qubit decoherence, internal cavity losses, and finite-duration pulses, in addition to transmission losses.
\end{abstract}

\maketitle

\section{Introduction}

The ability to scale up quantum computers is crucial for the long-term goal of performing computations fault-tolerantly. Many current proposals for scaling up take a modular approach where groups of qubits at the nodes of a network are linked by quantum photonic interconnects~\cite{reiserer2015cavity,vandersypen2017interfacing,awschalom2021development}. Recent experiments have shown significant progress, demonstrating entanglement distribution across nodes that are separated by tens of metres \cite{pompili2021realization,daiss2021quantum,storz2023loophole}. In a distributed architecture, the ability to perform long-range (inter-node) parity checks would also provide a clear path towards implementing quantum error correction~\cite{nickerson2013topological,nickerson2014freely}.

Many quantum error correcting codes, including the surface code~\cite{bravyi1998quantum,fowler2009high,fowler2012surface}, fall within a class of stabilizer codes~\cite{gottesman1997stabilizer} known as Calderbank-Shor-Steane (CSS) codes~\cite{calderbank1996good,steane1996multiple}. In CSS codes, errors can be detected by measuring the parities of groups of qubits, given by the eigenvalues of stabilizer operators of the form $X^{\otimes n}$ and $Z^{\otimes m}$, where $X$ and $Z$ are Pauli-$X$ and Pauli-$Z$ operators, respectively. Here, $n$ and $m$ are integers giving the weight of the stabilizer, equal to the number of qubits on which the operator acts nontrivially. A parity check can be performed using a sequence of entangling gates applied between an ancilla qubit and each of several code qubits. The ancilla qubit can then be measured to infer the parity of the code qubits. 

There are many strategies that can be used to implement two-qubit gates at long range. Given a source of distributed entanglement, these gates can be performed by teleportation \cite{gottesman1999demonstrating,huang2004experimental,jiang2007distributed,chou2018deterministic}. Alternatively, long-range two-qubit gates can be realized in solid state systems by coupling qubits via the real or virtual excitations of superconducting~\cite{majer2007coupling,bialczak2011fast,nigg2017superconducting,leroux2021superconducting,sung2021realization}, ferromagnetic~\cite{trifunovic2013long,fukami2021opportunities}, or normal-metal~\cite{trifunovic2012long} elements. 

Rather than mapping parity onto an ancilla qubit, parity information can more broadly be extracted by measuring any observable---not necessarily corresponding to a qubit observable---whose outcome depends on parity. In a circuit quantum electrodynamics (QED) architecture, for instance, this could be a shift in the resonance frequency of a microwave resonator dispersively coupled to multiple code qubits~\cite{tornberg2010high,lalumiere2010tunable,nigg2013stabilizer,divincenzo2013multi,riste2013deterministic,royer2018qubit}. 

In an architecture involving photonic interconnects, a natural way to connect qubits is using pulses of light. An inter-node parity check could then be performed via qubit-state-selective phase shifts $\alpha\rightarrow e^{i\phi}\alpha$ on a light pulse prepared in a coherent state $\ket{\alpha}$. Such state-selective phase shifts have been analyzed theoretically as a means of implementing quantum gates \cite{duan2004scalable,wang2005engineering}. In addition, small conditional phase shifts ($\phi<\pi$) generated from weak light-matter interactions have been theoretically considered as a resource for parity measurements \cite{yamaguchi2006quantum,myers2007stabilizer}. More recently, large phase shifts ($\phi\simeq\pi$) have been realized experimentally in both the optical~\cite{tiecke2014nanophotonic,hacker2019deterministic} and microwave~\cite{kono2018quantum,besse2018single,besse2020parity,wang2022flying} domains. These experiments demonstrate the enabling technology for parity checks based on ``flying-cat'' states given by superpositions of macroscopically distinct coherent states, $\ket{\mathrm{C}_\pm}\propto\ket{ \alpha}\pm \ket{-\alpha}$, associated with propagating pulses of optical or microwave light. In ancilla-based parity checks (involving gates on a common ancilla qubit), measurement errors and gate errors typically have independent physical origins and thus occur with independent probabilities. This is not the case for flying-cat parity checks, where these two error sources are correlated with the average number of photons in the coherent state, $\lvert\alpha\rvert^2$.  In particular, there is a trade-off between the rate of measurement errors (errors made in discriminating $\ket{\pm\alpha}$) and the rate of errors on code qubits due to backaction arising from photon loss. While the coherent-state distinguishability is controlled by a finite overlap $\langle \alpha\vert -\alpha\rangle$ that decreases with increasing $|\alpha|^2$, the rate of photon loss increases with increasing $|\alpha|^2$. 

In this paper, we quantify the performance of flying-cat parity checks in the presence of photon loss. To do this, we derive a superoperator describing the backaction of photon loss on code qubits during a weight-3 flying-cat parity check. From this superoperator, error patterns on code qubits together with their associated error probabilities can be identified. This knowledge can then be used to optimize $\alpha$ to minimize the total error probability. High quality weight-3 parity checks have applications in both universal quantum computing with surface codes~\cite{bravyi2012subsystem} and entanglement distribution in multi-node quantum networks. As a short-term demonstration of these ideas, we further show that weight-3 Pauli-$X$ and Pauli-$Z$ parity checks can be used to prepare a particular six-qubit entangled state.  This state can be used for three-party controlled quantum teleportation of an arbitrary two-qubit state. It also provides a natural testbed for flying-cat parity checks in early implementations, en route to the more ambitious goal of performing fault-tolerant quantum computation.

The layout of this paper is as follows. In Sec.~\ref{parity-meters}, we describe how qubit-state-selective phase shifts imprint parity information onto propagating light pulses. In Sec.~\ref{measurement}, we discuss the aforementioned trade-off between measurement errors and errors due to photon loss. Next, in Sec.~\ref{resource-states}, we illustrate how flying-cat parity checks can be used to generate entanglement in a multi-node network. In particular, we show how they can be used to prepare a six-qubit entangled state that provides a resource for three-party controlled quantum teleportation. Finally, in Sec.~\ref{feasibility}, we establish conditions for implementing flying-cat parity checks in a state-of-the-art circuit QED architecture, accounting for imperfections due to finite-bandwidth pulses, internal cavity loss, qubit decoherence, and transmission losses.

\section{Coherent states as parity meters}
\label{parity-meters}
\begin{figure}
    \centering
    \includegraphics[width=0.45\textwidth]{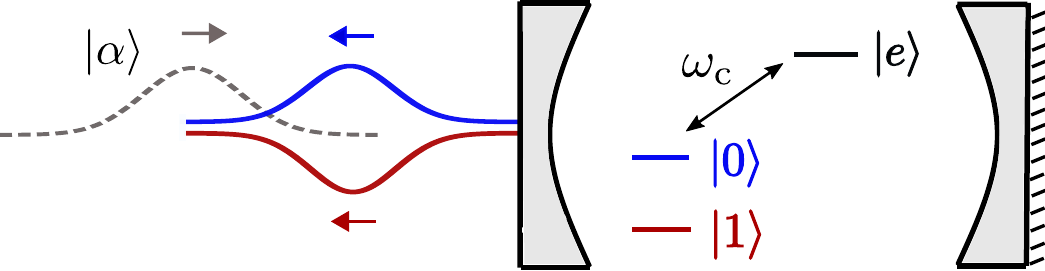}
    \caption{A qubit-state-selective phase shift on an incoming coherent state $\ket{\alpha}$ occupying some propagating quasimode of the electromagnetic field can be obtained by resonantly coupling a cavity mode having frequency $\omega_{\mathrm{c}}$ to the transition $\ket{0}\leftrightarrow\ket{e}$ involving qubit state $\ket{0}$ and some auxiliary level $\ket{e}$~\cite{duan2004scalable,wang2005engineering,besse2018single,hacker2019deterministic,besse2020parity}. }
    \label{fig:phase-shift}
\end{figure}

A parity measurement of two or more physical qubits can be performed by allowing a propagating light pulse in coherent state $\ket{\alpha}$ to interact with each qubit according to the entangling operation described by
\begin{align}
\begin{aligned}
    &\ket{0,\pm \alpha}\rightarrow \ket{0,\pm \alpha},\\
    &\ket{1,\pm \alpha}\rightarrow\ket{1,\mp \alpha},\label{entangle}
\end{aligned}
\end{align}
where $\ket{0}$ and $\ket{1}$ are the qubit Pauli-$Z$ eigenstates [with $Z\ket{s}=(-1)^s\ket{s}$ for $s=0,1$]. This operation can be realized by coupling a three-level system with states $\ket{1},\ket{0}$, and $\ket{e}$ to a single quantized cavity mode, such that the $\ket{0}\leftrightarrow\ket{e}$ transition is resonantly coupled to the cavity mode~\cite{duan2004scalable,wang2005engineering} (Fig.~\ref{fig:phase-shift}). An input coherent state $\ket{\alpha}$ occupying a propagating spatiotemporal mode (light pulse) that is resonant with the decoupled cavity frequency $\omega_{\mathrm{c}}$ will then acquire a different phase upon reflection from the cavity, depending on the qubit state. When the qubit is in state $\ket{1}$, the cavity frequency is unaffected by the presence of the qubit, and the input pulse is reflected with a $\pi$ phase shift~\cite{walls2008input}. However, when the qubit is in state $\ket{0}$, hybridization of the qubit and cavity suppresses the cavity density of states at $\omega_{\mathrm{c}}$, and the input pulse is reflected with no phase shift. This is of course not the only way of realizing the interaction given in Eq.~\eqref{entangle}. Another way to realize this scenario can be achieved in the dispersive regime (see Sec.~\ref{feasibility} below for a detailed discussion of this second strategy). 

After interacting with $n$ qubits in state $\ket{s_1s_2\dots s_n}$ ($s_i\in\{0,1\}$), $\ket{\alpha}$ will have acquired a total phase shift $(-1)^s=e^{is\pi}$, where $s=\sum_i s_i\:(\mathrm{mod}\:2)$ is the $Z$-basis parity of the $n$ qubits: 
\begin{equation}
    \ket{s_1s_2\dots s_n}\ket{\alpha}\rightarrow \ket{s_1s_2\dots s_n}\ket{e^{is\pi}\alpha}.
\end{equation}
The use of such an entangling interaction for continuous stabilization of a Bell state was analyzed theoretically in Ref.~\cite{kerckhoff2009physical} and demonstrated experimentally (with phase shifts $<\pi$ due to a small dispersive coupling) with superconducting qubits in Ref.~\cite{roch2014observation}.

Switching from $Z$-basis to $X$-basis parity checks simply requires that Hadamards be applied to the  qubits both before and after the entangling operation described in Eq.~\eqref{entangle}:
\begin{align}\label{hadamards}
    \begin{aligned}
        &\ket{+,\pm \alpha}\overset{H\;}{\rightarrow}\ket{0,\pm \alpha}\rightarrow\ket{0,\pm \alpha}\overset{H\;}{\rightarrow}\ket{+,\pm \alpha},\\
        &\ket{-,\pm \alpha}\overset{H\;}{\rightarrow}\ket{1,\pm \alpha}\rightarrow\ket{1,\mp \alpha}\overset{H\;}{\rightarrow}\ket{-,\mp \alpha},
    \end{aligned}
\end{align}
where here, $X\ket{\pm}=\pm \ket{\pm}$. Hence, for both $X$- and $Z$-basis parity checks, the parity of the code qubits can be inferred by measuring the phase of the coherent-state amplitude.

\subsection{Weight-3 checks are enough}

Many quantum error correcting codes are formulated as stabilizer codes~\cite{gottesman1997stabilizer} defined by a stabilizer group $\mathcal{S}$. The associated code space $C$ is then given by the span of the vectors $\ket{\psi}$ stabilized by $\mathcal{S}$, i.e., satisfying $S_j\ket{\psi}=\ket{\psi}$ for all stabilizers $S_j\in\mathcal{S}$. In the usual surface code~\cite{fowler2012surface}, for instance, correctable errors can be detected by measuring weight-4 stabilizers of the form $X^{\otimes 4}$ and $Z^{\otimes 4}$. 
\begin{figure*}
    \centering
    \includegraphics[width=\textwidth]{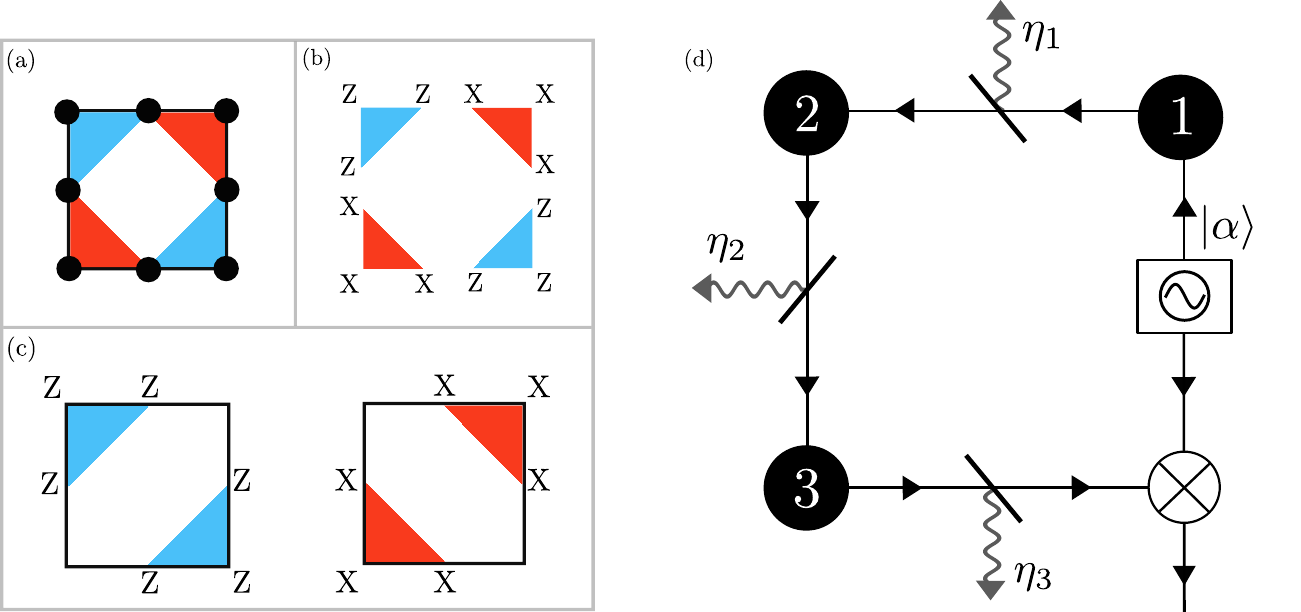}
    \caption{(a) The unit cell of the subsystem surface code~\cite{bravyi2012subsystem}. Physical qubits represented by solid black dots are located on the edges and vertices of a square lattice. (b) Gauge operators: Each stabilizer can be represented as the product of two weight-3 gauge operators. Gauge operators do not all mutually commute, but they do commute with all stabilizers and therefore preserve the code space. (c) Stabilizers of the subsystem surface code: Each unit cell hosts two stabilizers, one of the form $Z^{\otimes 6}$ (left) and one of the form $X^{\otimes 6}$ (right). (d) A light pulse is prepared in a coherent state $\ket{\alpha}$ and interacts sequentially with three qubits. The interaction of the light pulse with each qubit results in the entangling operation of Eq.~\eqref{entangle}. Photon loss in transit is modeled by a series of beam splitters having reflectivities $\eta_i$ ($i=1,2,3$). The  parity of the code qubits is inferred by performing a phase-sensitive measurement of the coherent state amplitude. This phase-sensitive measurement could be achieved through homodyne detection by interfering the light pulse with a reference signal and measuring the resulting intensity.}
    \label{fig:check}
\end{figure*}

A surface code can alternatively be operated as a subsystem code~\cite{bravyi2012subsystem} [Fig.~\ref{fig:check}(a-c)]. A subsystem code~\cite{poulin2005stabilizer,kribs2005unified,bombin2010topological} is defined by a non-Abelian gauge group $\mathcal{G}$ together with an Abelian stabilizer subgroup $\mathcal{S}\subseteq \mathcal{G}$. This construction allows the code space $C$ stabilized by $\mathcal{S}$ to be partitioned into subsystems, $C=A\otimes B$. Here, the subsystems $A$ and $B$ are defined by the action of gauge operators $g\in\mathcal{G}$, which may act nontrivially only on subsystem $B$: $g\ket{\psi_A}\otimes\ket{\psi_B}=\ket{\psi_A}\otimes\ket{\psi_B'}$ for $\ket{\psi_A}\in A,\,\ket{\psi_B},\ket{\psi_B'}\in B$. Errors acting on the codespace are then defined only up to an equivalence relation involving $\mathcal{G}$. In particular, error operators $E$ and $E'$ are equivalent in their action on encoded logical information (in subsystem $A$) if $E'=Eg$ for some $g\in\mathcal{G}$. Much of the appeal of subsystem codes comes from the fact that stabilizer eigenvalues can be inferred from measurements of lower-weight gauge operators. For the subsystem surface code~\cite{bravyi2012subsystem} [Fig.~\ref{fig:check}(a-c)], for instance, the eigenvalues of the weight-6 stabilizers $X^{\otimes 6}$ can be inferred by multiplying together the measurement outcomes of two weight-3 gauge operators of the form $X^{\otimes 3}$. Correctable errors in a surface code can therefore be detected using weight-3 parity checks only (as opposed to the usual weight-4 parity checks), with a fault-tolerance threshold of order 1\%~\cite{bravyi2012subsystem,higgott2021subsystem}.  Gates on logical qubits can be performed as in the usual surface code~\cite{fowler2012surface,bravyi2012subsystem}. 
 With this long-term application in mind, we consider error sources---photon loss and finite coherent-state distinguishability---that are specific to weight-3 flying-cat parity checks [Fig.~\ref{fig:check}(d)] and show how to minimize their total impact. Further error sources (finite-duration pulses, internal losses, and qubit decoherence) are assessed in Sec.~\ref{feasibility}, below.

An important advantage of the subsystem surface code is that errors on ancilla qubits used for syndrome readout propagate onto at most one other qubit, modulo gauge operators~\cite{bravyi2012subsystem}. The extra degree of freedom provided by the gauge group therefore eliminates the horizontal hook errors considered in Ref.~\cite{dennis2002topological} for the surface code, wherein ancilla-qubit errors propagate onto several qubits. 

\section{Trade-off between measurement errors and errors due to photon loss}
\label{measurement}

Parity checks performed using coherent states exhibit an unavoidable trade-off between error sources depending on the average number of photons per pulse, $\lvert \alpha\rvert^2$. Due to the non-orthogonality of $\ket{\pm \alpha}$, larger field amplitudes $\alpha$ yield better measurement fidelities. However, they also produce a stronger backaction on the code qubits in the presence of photon loss. In this section, we consider these two competing error sources: measurement errors and errors due to photon loss.

In a real-world implementation, there may be a number of other error sources affecting the quality of the operation required for parity checks [Eq.~\eqref{entangle}]. For example, the qubit-state-dependent conditional phase introduced may not be precisely $\pi$, phase noise or thermal light may corrupt the prepared coherent states, and photodetectors may suffer from amplifier noise, dark counts, etc. In this section, we assume that these implementation-dependent imperfections lead to a background error rate that can be controlled, leaving the effects of photon loss and imperfect coherent-state distinguishability as the dominant error sources. These effects are likely to be common to any implementation. We consider several additional sources of error in the specific context of a circuit-QED setup in Sec.~\ref{feasibility}.

\subsection{Photon-loss-induced backaction}

\begin{figure}
    \centering
    \includegraphics[width=0.44\textwidth]{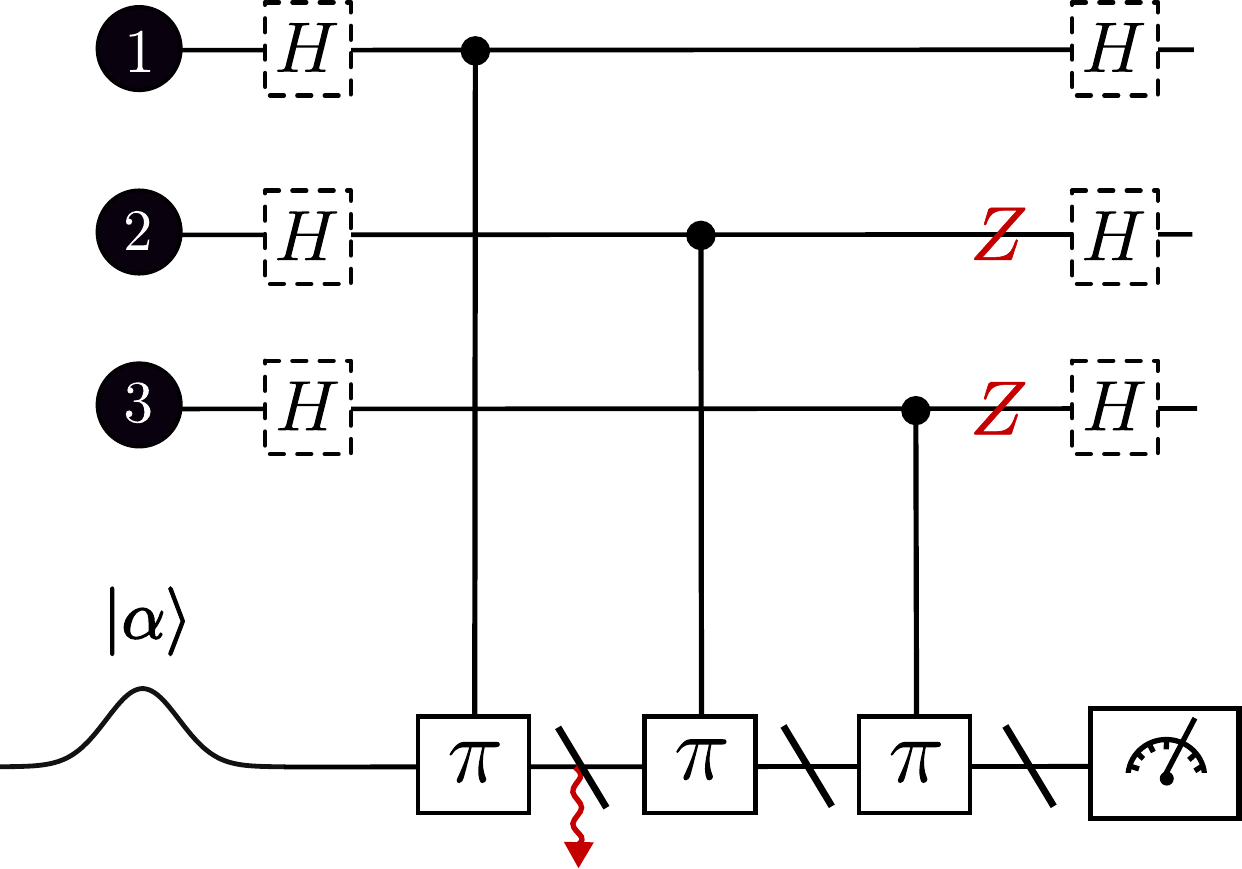}
    \caption{Horizontal hook error $Z_2Z_3$ due to photon loss during a parity check.}
    \label{fig:hook}
\end{figure}

In this subsection, we evaluate the backaction of photon loss on the state of the three code qubits involved in the parity check. Photon loss during a flying-cat parity check can lead to two-qubit correlated errors on the code qubits (horizontal hook errors, Fig.~\ref{fig:hook}). As is the case for parity measurements via noisy ancilla qubits~\cite{bravyi2012subsystem}, these horizontal hook errors are gauge-equivalent to single-qubit Pauli errors, and they can therefore be corrected as if they were single-qubit Pauli errors. This key advantage of the subsystem surface code is therefore retained for flying-cat parity checks. 

To perform a weight-3 parity check, a coherent state $\ket{\alpha}$ will  interact with three qubits. In what follows, it will be helpful to define the even and odd ($Z$-basis) parity subspaces given by
\begin{align}
\begin{aligned}
    &\mathcal{H}_+=\mathrm{span}\{\ket{000},\ket{110},\ket{011},\ket{101}\},\\
    &\mathcal{H}_-=\mathrm{span}\{\ket{001},\ket{010},\ket{100},\ket{111}\}.
\end{aligned}
\end{align}

We write the state $\ket{\Xi}$ of the three qubits prior to the parity check as
\begin{equation}
\ket{\Xi}=c_+\ket{\Xi_+}+c_-\ket{\Xi_-},\label{initial-state}
\end{equation}
where $\ket{\Xi_\pm}\in \mathcal{H}_\pm$, and where $\sum_{\sigma=\pm}\lvert c_\sigma\rvert^2=1$. In the absence of photon loss, the states $\ket{\Xi_\pm}$ are preserved by Eq.~\eqref{entangle}: $\ket{\Xi}\ket{\alpha}\rightarrow c_+\ket{\Xi_+}\ket{\alpha}+c_-\ket{\Xi_-}\ket{-\alpha}$. To perform an $X$-basis parity check, Hadamards are simply applied before/after the $Z$-basis parity check, as described in Eq.~\eqref{hadamards}. In that case, we take $\ket{\Xi}$ to describe the state of the three qubits after the first set of Hadamards.

In the presence of photon loss, the parity check has a backaction on the code qubits that leads to the introduction of errors beyond the background error rate.  We model the effect of photon loss to the environment via a beam splitter-type interaction~\cite{glancy2004transmission}
\begin{equation}
    R(\eta)=e^{\mathrm{arcsin}\sqrt{\eta}(a_\ell^\dagger a-\mathrm{h.c.})},\label{beamsplitter}
\end{equation}
where here, $a$ is an annihilation operator defined such that $a\ket{\alpha}=\alpha\ket{\alpha}$, $a_\ell$ is associated with an environmental loss mode into which photons may be scattered, and $\eta$ is the beam splitter reflectivity. Under this operation, $R^\dagger(\eta) a R(\eta)=\sqrt{1-\eta}a+\sqrt{\eta}a_\ell$. The amplitude $\alpha$ of $\ket{\alpha}$ is therefore reduced according to
\begin{equation}
    R(\eta)\ket{\alpha}\ket{0}_\ell=\ket{\sqrt{1-\eta}\alpha}\ket{\sqrt{\eta}\alpha}_\ell,
\end{equation}
where here, $\ket{\;}_\ell$ is the state of the loss mode. Although coherent-state amplitude may be lost to many different environmental modes in principle, these modes can be treated as a single effective mode since the state of the environment will eventually be traced over. To account for the effect of loss at different points in the parity-check operation, we imagine that the light pulse undergoes such a beam splitter interaction $R(\eta_i)$, $i=1,2,3$,  after every entangling operation [Fig.~\ref{fig:check}(d)]. Accounting for loss in this manner, the initial state $\ket{s_1s_2s_3}\ket{\alpha}\ket{0,0,0}_{\ell}$ of the qubits, coherent state, and environment (here, $\ket{0,0,0}_{\ell}$ denotes the vacuum state of the three loss modes) evolves into 
\begin{align}\label{bounce}
    \ket{s_1s_2s_3}\ket{(-1)^s\bar{\alpha} }\ket{(-1)^{s_1}\alpha_1,(-1)^{\mu}\alpha_2,(-1)^s\alpha_3}_{\ell},
\end{align}
where $s=s_1+s_2+s_3\:(\mathrm{mod}\:2)$ is the $Z$-basis parity of $\ket{s_1s_2s_3}$, $\mu=s_1+s_2\:(\mathrm{mod}\:2)$, and where 
\begin{align}
\begin{aligned}
    &\bar{\alpha}=\alpha\prod_{i=1}^3(1-\eta_i)^{1/2},\\
    &\alpha_i=\alpha\eta_i^{1/2}\prod_{j<i}(1-\eta_i)^{1/2}.\label{amplitudes}
\end{aligned}
\end{align}

The joint state $\rho$ of the qubits and electromagnetic field following the entangling operations (but prior to measurement) can then be obtained by tracing over the loss modes:
\begin{equation}
\rho=\sum_{\sigma,\sigma'=\pm} \rho_{\sigma,\sigma'}\ketbra{\sigma \bar{\alpha}}{\sigma'\bar{\alpha}},\label{pre-measurement}
\end{equation}
where $\rho_{\pm,\pm}\in\mathcal{H}_\pm$. For  $\ket{\Xi_+}=\sum_{i=0,3,5,6}c_i\ket{i}$, where here, $i$ is the base-10 value of $s_1s_2s_3$ in binary, $\rho_{+,+}$ takes the following form in the basis $\{\ket{0}=\ket{000},\ket{3}=\ket{011},\ket{5}=\ket{101},\ket{6}=\ket{110}\}$:
\begin{widetext}
\begin{equation}
    \frac{\rho_{+,+}}{\lvert c_+\rvert^2}=\begin{pmatrix}
        \lvert c_0\rvert^2 & c_0c_3^*e^{-2\lvert \alpha_2\rvert^2} & c_0c_5^*e^{-2(\lvert\alpha_1\rvert^2+\lvert\alpha_2\rvert^2)} & c_0c_6^*e^{-2\lvert\alpha_1\rvert^2} \\
        c_3c_0^*e^{-2\lvert\alpha_2\rvert^2} & \lvert c_3\rvert^2 & c_3c_5^*e^{-2\lvert \alpha_1\rvert^2} & c_3c_6^* e^{-2(\lvert\alpha_1\rvert^2+\lvert\alpha_2\rvert^2)} \\
        c_5c_0^*e^{-2(\lvert\alpha_1\rvert^2+\lvert\alpha_2\rvert^2)} & c_5c_3^* e^{-2\lvert\alpha_1\rvert^2} & \lvert c_5\rvert^2 & c_5c_6^*e^{-2\lvert\alpha_2\rvert^2}\\
        c_6c_0^* e^{-2\lvert\alpha_1\rvert^2} & c_6c_3^*e^{-2(\lvert\alpha_1\rvert^2+\lvert\alpha_2\rvert^2)} & c_6c_5^* e^{-2\lvert\alpha_2\rvert^2} & \lvert c_6\rvert^2
    \end{pmatrix}.\label{dephasing-due-to-loss}
\end{equation}
\end{widetext}
The block $\rho_{-,-}/\lvert c_-\rvert^2$ takes the same form as Eq.~\eqref{dephasing-due-to-loss} under the mapping $(0,3,5,6)\mapsto (1,2,4,7)$. The off-diagonal blocks $\rho_{\pm,\mp}\propto c_\pm c_\mp^*$ in Eq.~\eqref{pre-measurement} appear as a result of considering an initial state $\ket{\Xi}$ [Eq.~\eqref{initial-state}] of mixed parity and will be suppressed exponentially $\sim e^{-2\lvert\bar{\alpha}\rvert^2}$ in the final post-measurement state. 

In the absence of photon loss ($\eta_j=0$ for $j=1,2,3$), $\rho_{\sigma,\sigma'}=c_\sigma c_{\sigma'}^*\ketbra{\Xi_\sigma}{\Xi_{\sigma'}}$ and $\bar{\alpha}=\alpha$. With $\eta_j\neq 0$, however, $\rho_{\sigma,\sigma'}\neq c_\sigma c_{\sigma'}^*\ketbra{\Xi_\sigma}{\Xi_{\sigma'}}$ due to the photon-loss-induced backaction of the parity check. Hence, while the interaction with the electromagnetic field remains quantum non-demolition (QND) in the parity of $\ket{s_1s_2s_3}$ in the presence of photon loss, it is no longer QND in $\ket{\Xi_\pm}$ as desired. It may be verified that the backaction leading to this non-QND character can be described by the composition of two dephasing channels:
\begin{equation}
    \frac{\rho_{\pm,\pm}}{\lvert c_\pm\rvert^2}=(\mathcal{E}_{p_1,E_1}\circ\mathcal{E}_{p_2,E_2})(\ketbra{\Xi_\pm}),\label{channel}
\end{equation}
where $\mathcal{E}_{p,E}(\rho)=(1-p)\rho+p E\rho E^\dagger$, $E_1=Z_2Z_3$, $E_2=Z_3$, and where
\begin{align}\label{p1andp3}
    \begin{aligned}
    &p_1=\frac{1}{2}(1-e^{-2\lvert\alpha_1\rvert^2})\simeq \eta_1\lvert \alpha\rvert^2,\\
    &p_2=\frac{1}{2}(1-e^{-2\lvert\alpha_2\rvert^2})\simeq \eta_2\lvert\alpha\rvert^2.
    \end{aligned}
\end{align}
The approximations in Eq.~\eqref{p1andp3} are valid to leading order in $\eta_j\ll 1$.

The operators $E_{1,2}$ can be understood as follows: In the cat-state basis $\ket{\mathrm{C}_\pm}\propto \ket{\alpha}\pm \ket{-\alpha}$, Eq.~\eqref{entangle} describes a CZ gate for any nonzero $\alpha$. Since $\ket{\mathrm{C}_\pm}$ are eigenstates of the photon-number parity operator $e^{i\pi a^\dagger a}$, the loss of a photon constitutes an $X$ error, $\ket{\mathrm{C}_\pm}\to\ket{\mathrm{C}_\mp'}$, up to a change in coherent-state amplitude, $\ket{\alpha}\rightarrow\ket{\alpha'}$~\cite{cochrane1999macroscopically}. Hence, as a result of the CZ gates involved in the parity check, photon loss leads to $Z$ errors on the code qubits. (Generally speaking, an $X$ error on qubit $a$, propagated through a CZ acting on qubits $a$ and $b$, leads to a $Z$ error on qubit $b$.) These $Z$ errors appear in the combination $E_1=Z_2Z_3$ for photon loss occurring between qubits 1 and 2  (Fig.~\ref{fig:hook}), and as $E_2=Z_3$ for photon loss occurring between qubits 2 and 3.

In a surface code where stabilizer parity checks are performed using propagating pulses of light, photon loss would therefore lead to correlated (horizontal hook) errors on code qubits. However, due to the gauge degree of freedom introduced by the gauge group, the horizontal hook error $Z_2Z_3$ introduced by the parity check [cf.~Eq.~\eqref{channel}] is equivalent to the single-qubit error $Z_1$~\cite{bravyi1998quantum}. This is because $Z_2Z_3(Z_1Z_2Z_3)=Z_1$, where $Z_1Z_2Z_3\in\mathcal{G}$. Since $X$-basis parity checks differ only by some additional Hadamards applied to $\rho$ [Eq.~\eqref{pre-measurement}], the effects of photon loss during $X$-basis parity checks are instead equivalent, up to a gauge operator, to single-qubit $X_1$ and $X_3$ errors occurring with the same probabilities (Table \ref{1qerrors}). These errors are not detected by the parity check in which they are introduced (since $Z$-basis parity checks only detect $X$ errors and vice versa), but may be detected in the next parity check of the other basis. 
\begin{table}
\centering
\resizebox{0.44\textwidth}{!}{
\begin{tabular}{c||c|c}    \toprule
Operator  & Error $E$ introduced& Prob($E$)\\ being measured &by photon loss& \\ \toprule
$Z^{\otimes 3}$ & $Z_1$ & $p_1$\\
& $Z_3$ & $p_2$\\
$X^{\otimes 3}$ & $X_1$ & $p_1$\\
& $X_3$ & $p_2$\\\bottomrule
\end{tabular}}
\caption{Errors arising due to photon loss during a $Z^{\otimes 3}$ gauge-operator parity check are equivalent (up to a gauge transformation) to $Z_1$ or $Z_3$ errors arising with probabilities $p_1$ and $p_2$, respectively. During an $X^{\otimes 3}$ parity check, these errors are instead equivalent to $X_1$ or $X_3$ errors.   }
\label{1qerrors}
\end{table}

\subsection{Measurement errors}

Determining the parity of two or more qubits using this approach requires a phase-sensitive measurement of the electromagnetic field (to infer the sign of $\alpha$). Such a measurement can be implemented via homodyne detection~\cite{tyc2004operational}. In homodyne detection, a signal field is first mixed on a 50:50 beam splitter with a local oscillator in coherent state $\ket{\beta}$, and the output intensity is then measured, giving information about the phase of the signal relative to the local oscillator. In this case, the signal field is the propagating light pulse entangled with the state of the qubits [Eq.~\eqref{pre-measurement}]. Without loss of generality, we assume that $\alpha$ is real and positive, $\alpha\in\mathbb{R}^+$. With a strong local oscillator ($\beta\in\mathbb{R}\gg \alpha)$, homodyne detection implements a projective measurement $\ketbra{x}$ onto the $x$ eigenbasis~\cite{tyc2004operational}, where here, $\ket{x}$ is an eigenstate of $\hat{x}=(a^\dagger+a)/\sqrt{2}$. The sign $\pm =\mathrm{sign}(x)$ of the measured displacement can then be taken as an inference of even ($+$) or odd ($-$) parity for the state of the qubits. In the event that $\alpha$ is complex, the measured quadrature can be adjusted by introducing a phase on the local oscillator: $\beta\rightarrow \beta e^{i(\varphi-\pi)}$. In that case, the projection is instead onto an eigenstate of $\hat{x}_\varphi=\hat{x}\cos{\varphi}+\hat{p}\sin{\varphi}$, where $[\hat{x},\hat{p}]=i$~\cite{tyc2004operational}.

Given $\rho$ [Eq.~\eqref{pre-measurement}], the state $\rho_x$ of the code qubits conditioned on the measurement outcome $x$ is then given by
\begin{equation}
    \rho_x=\frac{\mathrm{Tr}_{\mathrm{EM}}\{\ketbra{x}{x}\rho\ketbra{x}{x}\}}{p(x)},\label{post-meas}
\end{equation}
where $p(x)=\mathrm{Tr}\{\ketbra{x}\rho\}$, and where $\mathrm{Tr}_{\mathrm{EM}}\{\dotsm\}$ denotes a partial trace over the state of the electromagnetic field. This quantity can be evaluated straightforwardly using the representation $\langle x\vert\bar{\alpha}\rangle=\pi^{-1/4}e^{-\frac{1}{2}(x-\sqrt{2}\bar{\alpha})^2}$ of $\ket{\bar{\alpha}}$. For a thresholded decision, where $x>0$ ($x<0$) is taken to indicate even parity (odd parity), the post-measurement state for an inference of even parity is given by
\begin{equation}
    \varrho_+=\int_0^\infty dx\:p(x)\rho_x,
\end{equation}
with a similar expression for $\varrho_{-}$ involving an integral from $-\infty$ to 0. The joint probability $P(``\pm",\mp)$ of the state being odd/even ($\mp$) for an inference of even/odd parity ($``\pm"$) is 
\begin{equation}
    P(``\pm",\mp)=\mathrm{Tr}\{\hat{\Pi}_\mp\varrho_\pm\}=\frac{\lvert c_\mp\rvert^2}{2}\mathrm{erfc}(\sqrt{2}\bar{\alpha}),
\end{equation}
where $\hat{\Pi}_\pm$ is a projector onto $\mathcal{H}_\pm$ and $\mathrm{erfc}(x)$ is the complementary error function. The error $p_\mathrm{M}$ associated with the measurement is then given by
\begin{align}\label{measurement-error}
\begin{aligned}
    p_{\mathrm{M}}=P(``+",-)+P(``-",+)=\frac{1}{2}\mathrm{erfc}(\sqrt{2}\bar{\alpha}).
\end{aligned}
\end{align}

\begin{figure}
    \centering
    \includegraphics[width=0.45\textwidth]{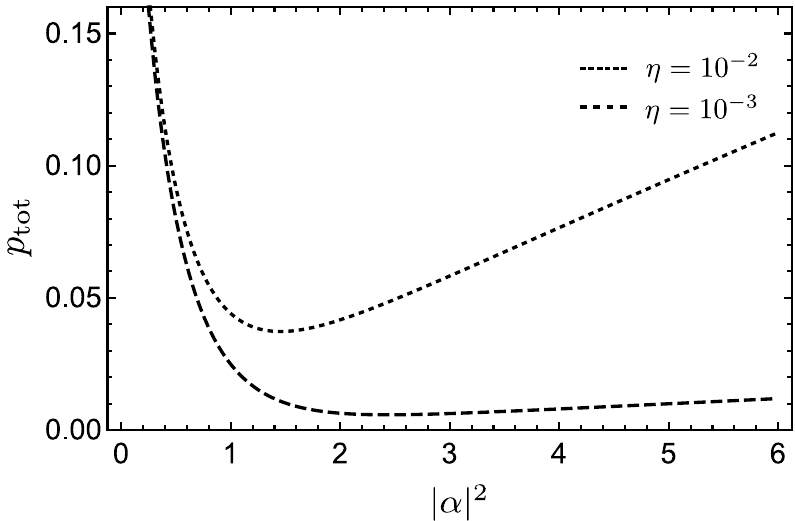}
    \caption{The total error $p_{\mathrm{tot}}$ [Eq.~\eqref{total-error}] is given by the sum of competing terms: measurement errors, which decrease with $\alpha$, and errors due to photon loss, which increase with $\alpha$. For simplicity, we assume that $\eta_i=\eta$ for all $i$.}
    \label{fig:tradeoff}
\end{figure}

The total probability of an error occurring as part of a parity check---either an incorrect parity inference or a code-qubit error---is then (neglecting corrections of order $p_i p_j$ for $p_i\ll 1$):
\begin{equation}
    p_{\mathrm{tot}}=p_{\mathrm{M}}+p_1+p_2\simeq \frac{1}{2}\frac{e^{-2\alpha^2}}{\alpha\sqrt{2\pi}}+\frac{1}{2}\sum_{j=1,2} \eta_j\lvert\alpha\rvert^2.\label{total-error}
\end{equation}
In the second, approximate equality, we have neglected subleading corrections in $\eta_{j}|\alpha|^2<1$ and approximated the error function by its asymptotic form for $|\alpha|>1$. Since errors introduced by photon loss occur at a rate that increases with $\alpha$ ($\sim \eta_j\lvert\alpha\rvert^2$), while measurement errors are suppressed for large $\alpha$ ($\sim \alpha^{-1}e^{-2\alpha^2}$), there is an inherent trade-off between the two types of error as a function of coherent-state amplitude. Their total impact can therefore be minimized by optimizing $\alpha$ (Fig.~\ref{fig:tradeoff}).

In light of the nonlinear (exponential) suppression of $p_\mathrm{M}$ with $|\alpha|^2$, it is natural to consider whether repeating the measurement can lead to a smaller error. In this case, a single-shot measurement with a coherent state of amplitude $\alpha=\sqrt{N}\alpha_0$ leads to the same measurement error $p_\mathrm{M}$ as a sequence of $N$ measurements performed with smaller coherent state amplitudes $\alpha_0$. However, this equivalent performance can only be reached if a soft decision is made, where each measurement outcome $x_k$ $(k=1,2,\ldots,N)$ is assigned a confidence $p(\pm|x_k)=\mathrm{Tr}\{\hat{\Pi}_\pm\rho_{x_k}\}$, and where the individual outcomes are correlated in determining the most likely parity. If instead, each of the $N$ measurements is thresholded independently leading to a majority-vote decision, the error is less favorable, $p_\mathrm{M,th}\sim \sqrt{p_\mathrm{M}}$ (up to logarithmic corrections for large $N$) \cite{d2014soft}. The large discrepancy between the error for a thresholded (hard) decision and the soft decision is a feature of the Gaussian distribution of measurement outcomes $x_k$ \cite{d2014soft,xue2020repetitive}, a model that is accurately realized for measurement of a coherent state quadrature. Exploiting this soft-decision advantage in repeated measurements with weak pulses may prove useful if the amplitude of each coherent state is limited for technical reasons (nonlinearity, unwanted qubit excitation, amplifier noise, etc.).

\section{Entangled resource states}
\label{resource-states}

In addition to enabling quantum error correction, flying-cat parity checks could also be used to distribute entanglement in a quantum network, with applications in both quantum computing and quantum communication. In this section, we discuss the concrete example of generating three-qubit Greenberger-Horne-Zeilinger (GHZ) states. GHZ states have applications in precision sensing and multi-party quantum communication. They could also be used as the central resource for weight-3, ancilla-based inter-node parity checks using the approach of Refs.~\onlinecite{nickerson2013topological,nickerson2014freely} (an alternative to the strategy for direct parity checks described above). We further show how to create a particular six-qubit entangled ``tetrahedron'' state using only $X$- and $Z$-basis weight-3 parity checks and single-qubit gates. This state can be used as a resource for three-party controlled quantum teleportation of an arbitrary two-qubit state. Its preparation also provides a convenient setting in which to benchmark the performance of the parity checks themselves.

\subsection{A three-qubit GHZ state}

An alternative, established approach to long-range parity checks involves the generation and consumption of four-qubit GHZ states~\cite{nickerson2013topological,nickerson2014freely}. In this setup, each node houses a code qubit and (at least) one ancilla. To perform a parity check, four ancilla qubits located at four different nodes are prepared in a GHZ state. The parity of the code qubits located at these nodes can then be inferred by applying controlled-NOT gates between the code qubit and the ancilla located at each node, then measuring all ancillae. The distributed GHZ state in this proposal could be generated, in principle, via flying-cat parity checks. Although the original proposal of Refs.~\onlinecite{nickerson2013topological,nickerson2014freely} focused on weight-4 parity checks, weight-3 checks are sufficient to operate a subsystem surface code, as described above, so in this section we consider the preparation of three-qubit GHZ states. The ancilla-based approach of Refs.~\onlinecite{nickerson2013topological,nickerson2014freely} may be preferable over a direct flying-cat parity check in the case where the photon loss rate is high, leading to a high probability of errors being introduced on the code qubits. The low-fidelity GHZ states produced in the presence of high photon loss could then be purified ``offline'' prior to being consumed as part of a parity check, as considered in Refs.~\cite{nickerson2013topological,nickerson2014freely}.

A simple measurement-based strategy for three-qubit GHZ-state preparation begins with preparation of the three qubits in $\ket{+++}$. Measuring $Z_1Z_2$ and $Z_2Z_3$ projects the qubits into an entangled state, which can then be transformed into $\ket{\mathrm{GHZ}}\propto \ket{000}+\ket{111}$ with an $X$-gate correction on the appropriate qubit. For a measurement outcome $Z_1Z_2=-1$, this is qubit 1; for $Z_2Z_3=-1$, it is qubit 3; if both outcomes are $-1$, it is qubit 2. However, as described above in Sec.~\ref{measurement}, a $Z$-basis flying-cat parity check may introduce $Z$ errors on the qubits. For a measurement of $Z_iZ_j$, these are single-qubit $Z_j$ errors occurring with probability $p_{ij}=(1-e^{-2\eta_{ij}\lvert\alpha\rvert^2})/2$, where here, $\eta_{ij}$ is the beam splitter reflectivity quantifying the strength of the losses incurred while traveling between qubits $i$ and $j$. Measurement errors occurring with probability $q_{ij}=\mathrm{erfc}[\sqrt{2}(1-\eta_{ij})\alpha]/2$ will also result in the wrong correction operator being applied. In the presence of photon loss and accounting for measurement errors, the two parity checks and correction step described above will lead to preparation of the following mixed state:
\begin{equation}
    \sigma=(1- p)\ketbra{\mathrm{GHZ}}+\sum_ip_i\ketbra{i_\perp},
\end{equation}
where $p=\sum_i p_i=p_{12}+p_{23}-p_{12}p_{23}+q_{12}+q_{23}+q_{12}q_{23}$, and where $\langle \mathrm{GHZ}\vert i_\perp\rangle=0$ for all $i$. 

Since $\ket{\mathrm{GHZ}}$ can be described as the simultaneous $+1$ eigenstate of a set of CSS stabilizer generators---namely, $Z_1Z_2$, $Z_2Z_3$, and $X_1X_2X_3$---several noisy copies of $\sigma$ distributed across three nodes (``Alice'', ``Bob'', and ``Charlie'') could be purified using any CSS error-correcting code, e.g., $C_4$~\cite{glancy2006entanglement}. In the case of $C_4$, this protocol would require that Alice, Bob, and Charlie share four copies of $\sigma$. They would each measure the stabilizers of $C_4$ on their four qubits, and then use local operations and classical communication to produce a less noisy, encoded version of $\sigma$~\cite{glancy2006entanglement}. Since the six-qubit ``tetrahedron'' state discussed in the following subsection also has a description in terms of a set of CSS stabilizer generators, the same protocol~\cite{glancy2006entanglement} could be used to purify this state as well.

\subsection{A six-qubit entangled state}
\begin{figure}
    \centering
    \vspace{4ex}
    \includegraphics[width=0.47\textwidth]{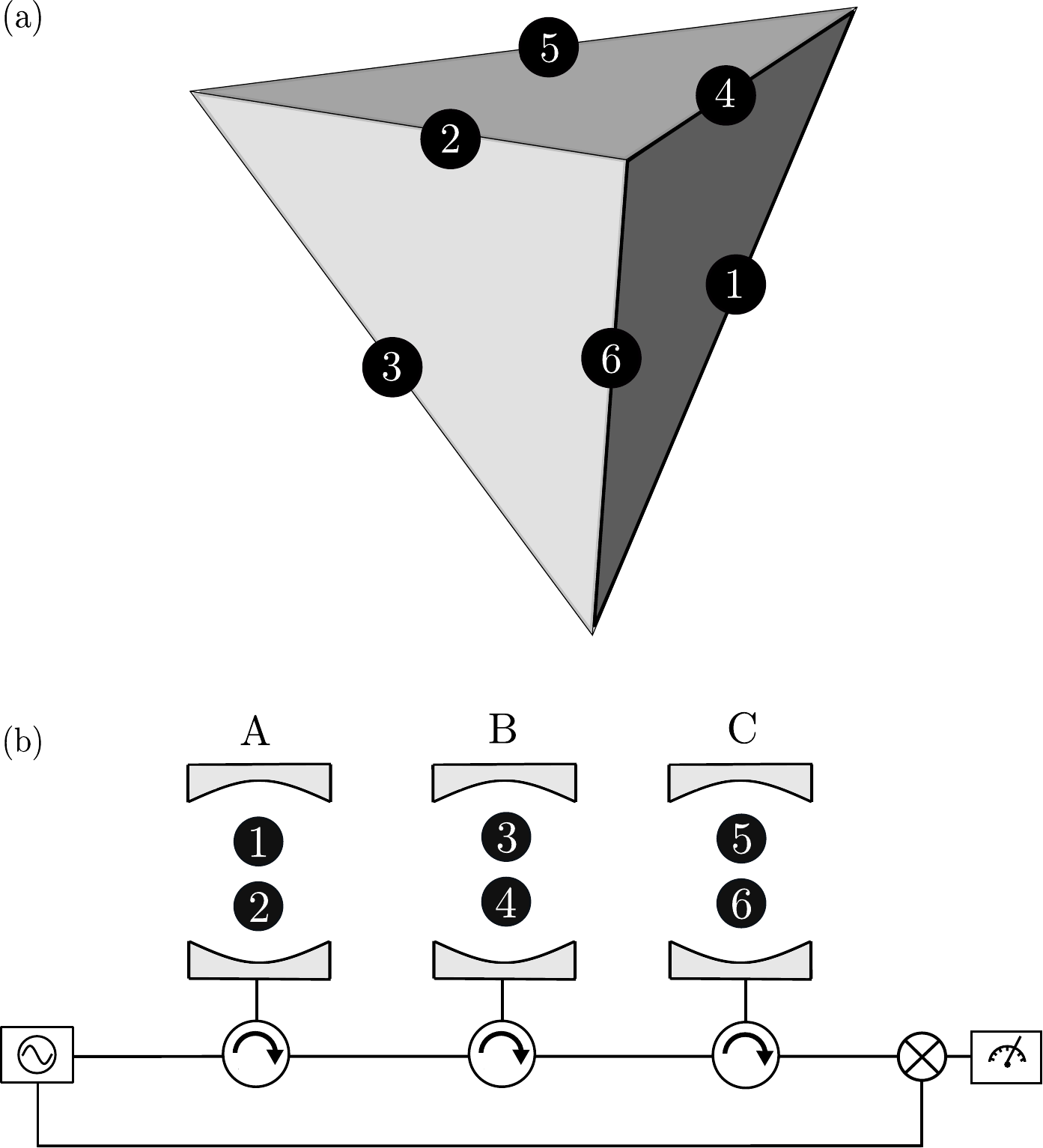}
    \caption{(a) Six qubits can be associated with the six edges of a tetrahedron. Each of the stabilizers in Eq.~\eqref{stabilizers} is a product of three Pauli-$Z$ or Pauli-$X$ operators acting on qubits that neighbor a common face ($Z$) or vertex ($X$). (b) A potential real-space setup for applications discussed in the main text: Each node ($A,B,C$) contains a pair of qubits whose associated tetrahedron edges have no face nor vertex in common. Every parity check required for preparation of $\ket{\mathrm{T}}$ [Eq.~\eqref{stabilizers}] involves only one qubit at each node. For nodes $(A,B,C)$ implemented via the cavity-QED realization shown in Fig.~\ref{fig:phase-shift}, the three qubits not being measured during a given parity check can be made inactive by detuning them from their respective cavities.}
    \label{fig:tetrahedron}
\end{figure}

In this section, we show how weight-3 parity checks can be used to prepare a ``tetrahedron'' state---a six-qubit entangled state where each qubit can be associated with one edge of a tetrahedron [Fig.~\ref{fig:tetrahedron}(a)]. This state provides a source of shared randomness with potential applications for three-party quantum key distribution. A tetrahedron state can also be used as a resource for controlled quantum teleportation of an arbitrary two-qubit state. Independent of these potential applications, the tetrahedron state provides a natural testbed for benchmarking the performance of weight-3 flying-cat parity checks (since its preparation requires both $X$- and $Z$-basis parity checks).

The tetrahedron state spans the codespace of a (trivial) CSS code admitting a set of stabilizer generators that can be associated with the faces and vertices of a tetrahedron [Fig.~\ref{fig:tetrahedron}(a)]:
\begin{align}
\begin{aligned}\label{stabilizers}
    &S_1=Z_1Z_3Z_5,\quad\quad S_4=X_1X_3X_6,\\
    &S_2=Z_1Z_4Z_6,\quad\quad S_5=X_1X_4X_5,\\
    &S_3=Z_2Z_4Z_5,\quad\quad S_6=X_2X_3X_5.\
\end{aligned}
\end{align}
It may be verified that the stabilizers associated with the remaining face and vertex are simply given by $S_1S_2S_3$ and $S_4S_5S_6$. 

The stabilizer eigenvector relations $S_i\ket{\mathrm{T}}=\ket{\mathrm{T}}$, $i=1,2,\dots,6$, admit one common eigenstate, the tetrahedron state:
\begin{align}
\begin{aligned}
    \ket{\mathrm{T}}=\frac{1}{2}
    \sum_{\beta=\Phi^\pm,\Psi^\pm}\ket{\beta}_{12}\ket{\beta}_{34}\ket{\beta}_{56},
\end{aligned}\label{memory}
\end{align}
where $\ket{\Phi^\pm}=(\ket{00}\pm\ket{11})/\sqrt{2}$ and $\ket{\Psi^\pm}=(\ket{01}\pm\ket{10})/\sqrt{2}$. If node $A$ houses qubits 1 and 2, node $B$ qubits 3 and 4, and node $C$ qubits 5 and 6 [Fig.~\ref{fig:tetrahedron}(b)], then the outcomes of Bell measurements at nodes $A$-$C$ are uniformly distributed across Bell states but perfectly correlated across nodes. A tetrahedron state can therefore provide Alice, Bob, and Charlie with two bits of shared randomness, which could be used to build a shared private key for cryptography.

With the qubits partitioned in the manner described above, every stabilizer $S_i$ in Eq.~\eqref{stabilizers} involves one qubit at each node. Preparation of $\ket{\mathrm{T}}$ can be achieved in two steps, starting from a random initial state. In the first step, all $S_i$ are measured using flying-cat parity checks. At this point, each stabilizer measurement returns a random outcome, $\sigma_i=\pm 1$. In the second step, the $-1$ outcomes are treated like an error syndrome and the appropriate correction operator (identified in the manner described below) is applied, transforming the post-measurement state into the target state $\ket{\mathrm{T}}$.

Since $\ket{\mathrm{T}}$ [Eq.~\eqref{memory}] is the non-degenerate ground state of $-\sum_i S_i$, the code defined by $S_i$ is trivial, reflecting the trivial genus-0 topology of the tetrahedron.  Despite the absence of logical qubits (which, in a toric-code construction~\cite{kitaev2003fault}, requires that $-\sum_iS_i$ have degenerate ground states), the code nonetheless serves as a quantum memory for $\ket{\mathrm{T}}$: Given full syndrome information, consisting of the six eigenvalues $\sigma_i$ of $S_i$, any number of $X$ and $Z$ errors can be corrected using a simple syndrome-decoding algorithm. The procedure is identical for $X$ and $Z$ errors, so without loss of generality, we give the decoding procedure for $X$ errors:
\begin{enumerate}[itemsep=-1ex]
    \item Measure the $Z$-basis stabilizers $S_{1-3}$. If the syndrome is $(1,1,1)$, then no correction is required. All other errors can be divided into two classes:
    \item  Class I errors: If syndrome $j$ in Table \ref{syndrome} is obtained, then apply an $X$ gate to qubit $j$.
    \item Class II errors: If the syndrome consists entirely of $-1$ outcomes, then apply $X$ gates to \textit{any} pair $(1,2)$, $(3,4)$, or $(5,6)$ of qubits having no face nor vertex in common.
\end{enumerate}
\begin{table}
\vspace{3ex}
\resizebox{0.42\textwidth}{!}{
\begin{tabular}{c||ccc|ccc}    \toprule $j\;\;$ & $\;\;\sigma_1\;\;$ & $\;\;\sigma_2\;\;$ & $\;\;\sigma_3\;\;$ & $\;\;\sigma_4\;\;$ & $\;\;\sigma_5\;\;$ & $\;\;\sigma_6\;\;$ \\\toprule
$1\;\;$ & $-1$ & $-1$ & $+1$  & $-1$ & $-1$ & $+1$  \\
$2\;\;$ & $+1$ & $+1$ & $-1$  & $+1$ & $+1$ & $-1$  \\ 
$3\;\;$ & $-1$ & $+1$ & $+1$  & $-1$ & $+1$ & $-1$  \\
$4\;\;$ &  $+1$ & $-1$ & $-1$ & $+1$ & $-1$ & $+1$ \\
$5\;\;$ & $-1$ & $+1$ & $-1$ & $+1$ & $-1$ & $-1$  \\
$6\;\;$ &  $+1$ & $-1$ & $+1$ & $-1$ & $+1$ & $+1$  \\
\bottomrule 
 \hline 
\end{tabular}}
\caption{Syndrome data for decoding Class I errors.   }
\label{syndrome}
\end{table}
Any number of $X$ errors can be corrected in this way: Single-qubit errors are reversed, while multi-qubit $X$ errors \textit{corrected as single-qubit $X$ errors} are transformed either into $S_{4-6}$ or into products thereof, all of which act trivially on $\ket{\mathrm{T}}$. As an example, consider the scenario where qubits 1 and 3 have undergone bitflip errors. The syndrome $(\sigma_1,\sigma_2,\sigma_3)=(1,-1,1)$ dictates that an $X$ gate be applied to qubit 6 (Table \ref{syndrome}). This operation transforms the state $X_1X_3\ket{\mathrm{T}}$ into $S_4\ket{\mathrm{T}}=\ket{\mathrm{T}}$.  For an example of a Class II error, consider the scenario where both qubits 1 and 2 have undergone $X$ errors. In this case, $(\sigma_1,\sigma_2,\sigma_3)=(-1,-1,-1)$, requiring that $X$ gates be applied to any pair $(1,2)$, $(3,4)$, or $(5,6)$ of qubits. If $(1,2)$ is selected, then the errors are reversed. If $(3,4)$ is selected instead, then the correction operator  $X_3X_4$ transforms $X_1X_2\ket{\mathrm{T}}$ into $S_5S_6\ket{\mathrm{T}}=\ket{\mathrm{T}}$. 

Multipartite entanglement of $\ket{\mathrm{T}}$ can be detected by measuring the entanglement witness~\cite{terhal2002detecting,bourennane2004experimental} 
\begin{equation}
    \mathcal{W}=\alpha_{\ket{\mathrm{T}}}\mathbbm{1}-\ketbra{\mathrm{T}},\label{witness-def}
\end{equation}
where here, $\alpha_{\ket{\mathrm{T}}}$ is the largest Schmidt coefficient across bipartitions of $\ket{\mathrm{T}}$. This construction is designed to have a non-negative expectation value $\langle\mathcal{W}\rangle\geq 0$ for all biseparable states. A negative expectation value therefore signals detection of genuine multipartite entanglement.

For states (like $\ket{\mathrm{T}}$) having a description in terms of a set of CSS stabilizer generators, it is known that the maximal Schmidt coefficient across bipartitions is $\alpha_{\ket{\mathrm{T}}}=1/2$, and that the witness $\mathcal{W}$ can be expressed in terms of the CSS stabilizer generators~\cite{guhne2009entanglement}: 
\begin{equation}
    \mathcal{W}=\frac{3}{2}\mathbbm{1}-\prod_{i=1}^3\frac{\mathbbm{1}+S_i}{2}-\prod_{i=4}^6\frac{\mathbbm{1}+S_i}{2}.\label{witness}
\end{equation}
The expectation value $\langle \mathcal{W}\rangle$ can be estimated using two measurement settings only: one where all qubits are measured in the $Z$ basis, and one where all qubits are measured in the $X$ basis. The fidelity $F=\langle \mathrm{T}\lvert \rho\rvert\mathrm{T}\rangle=1/2-\langle\mathcal{W}\rangle$ of some noisy state $\rho$ relative to $\ket{\mathrm{T}}$ can therefore be estimated using single-qubit measurements, independent of the parity checks that were used to create the state. This provides an avenue for benchmarking the quality of the weight-3 flying-cat parity checks used to prepare $\ket{\mathrm{T}}$.

When a complete set of parity checks is performed starting from some initial state, a set of six eigenvalues $\left\{\sigma_i\right\}$ is found. In the absence of measurement errors or errors due to photon loss, the operation required to transform the initial state into $\ket{\mathrm{T}}$ can be correctly identified based on these eigenvalues. A measurement error, however, will lead to the wrong operation being applied, resulting in the creation of a state orthogonal to $\ket{\mathrm{T}}$. Errors on physical qubits can also be introduced by photon loss during a parity check. This loss-induced backaction takes the form of stochastic $Z$ errors ($X$ errors) during measurement of $Z$-basis ($X$-basis) parity checks. Such errors could be corrected in subsequent error-correction cycles using the decoding procedure explained above. However, any errors introduced by the \textit{last} set of parity checks (prior to any measurements occurring at nodes $A$-$C$) remain uncorrected. If, for instance, this last round of checks consists of $Z$-basis parity checks, then $\ket{\mathrm{T}}$ may be affected by uncorrected $Z$ errors, which also produce states orthogonal to $\ket{\mathrm{T}}$. To leading order in $p_{1,2}$, and neglecting corrections $O(p_{1,2}p_{\mathrm{M}})$, we then have
\begin{equation}
    \langle\mathcal{W}\rangle\simeq -\frac{1}{2}+[1-(1-p_{\mathrm{M}})^6]+3(p_1+p_2).
\end{equation}
Here, the term $[1-(1-p_{\mathrm{M}})^6]$ gives the probability of making at least one measurement error across the six stabilizer measurements, while the term $3(p_1+p_2)$ gives the probability of photon loss introducing an error during the last set of parity checks (consisting of either three $X$-basis or three $Z$-basis checks). Notably, in the limit $\alpha\rightarrow 0$, we have $\langle\mathcal{W}\rangle=1/2-1/2^6$, giving $F=1/2^6$. This is the fidelity expected from randomly guessing the six stabilizer outcomes and applying a correction based on these guesses.

\begin{figure}
    \centering
    \includegraphics[width=0.45\textwidth]{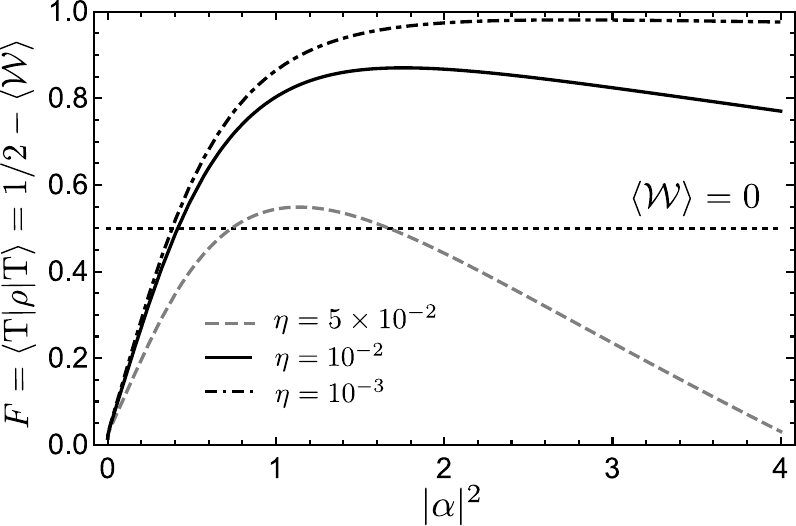}
    \caption{The preparation fidelity $F=\langle \mathrm{T}\lvert \rho\rvert\mathrm{T}\rangle$ accounting for measurement errors and errors due to photon loss. }
    \label{fig:witness}
\end{figure}

\subsubsection{Controlled quantum teleportation}

The tetrahedron state can be used as a resource state for three-party controlled quantum teleportation. Controlled quantum teleportation of a single-qubit state using a three-qubit GHZ state was introduced in Ref.~\cite{karlsson1998quantum}, and since then, many protocols extending the scheme to multiqubit states using various resource states have been proposed~\cite{yang2004efficient,yang2005scheme,deng2005multiparty,li2006efficient,choudhury2009quantum,kumar2020experimental}. The premise underlying this protocol is that Alice and Bob should be able to teleport quantum states with unit fidelity only with Charlie's cooperation. Here, the fidelity of a teleportation protocol is given by the fidelity of the state reconstructed by Bob relative to Alice's original state. In Ref.~\cite{li2014control}, Li and Ghose introduced the notion of control power to quantify Charlie's influence over the fidelity of the teleportation protocol. The control power is equal to $1-\bar{F}$ for an average teleportation fidelity $\bar{F}$. For teleportation of a two-qubit state, Charlie's control power must be $1-\bar{F}\ge 3/5$ in order to prevent Alice and Bob from benefiting from their shared entanglement~\cite{li2015analysis}. Not all entangled states meet this criterion, but we now show that the state $\ket{\mathrm{T}}$ does.

We imagine that Alice, Bob, and Charlie share one copy of $\ket{\mathrm{T}}$, with each party holding two qubits according to the partitioning shown in Fig.~\ref{fig:tetrahedron}(b). Alice has two additional qubits $A_1,A_2$ in state
$\ket{\varphi}=a\ket{\Phi^+}+b\ket{\Psi^+}+c\ket{\Psi^-}+d\ket{\Phi^-}$, which she wants to teleport to Bob. She performs Bell measurements on pairs of qubits $(A_1,1)$ and $(A_2,2)$ and transmits the results to Bob, who applies the correction operators given in Table \ref{correction-ops}. Once these correction operators have been applied, Bob and Charlie share one of four possible states:
\begin{align} \label{bob-charlie}
\begin{aligned}
&a\ket{\Phi^+,\Phi^+}+b\ket{\Psi^+,\Psi^+}+c\ket{\Psi^-,\Psi^-}+d\ket{\Phi^-,\Phi^-},\\ 
& a\ket{\Phi^+,\Psi^+}+b\ket{\Psi^+,\Phi^+}+c\ket{\Psi^-,\Phi^-}+d\ket{\Phi^-,\Psi^-},\\
&a\ket{\Phi^+,\Psi^-}+b\ket{\Psi^+,\Phi^-}+c\ket{\Psi^-,\Phi^+}+d\ket{\Phi^-,\Psi^+},\\
&a\ket{\Phi^+,\Phi^-}+b\ket{\Psi^+,\Psi^-}+c\ket{\Psi^-,\Psi^+}+d\ket{\Phi^-,\Phi^+}.
\end{aligned}
\end{align}
\begin{table}
\vspace{3ex}
\resizebox{0.44\textwidth}{!}{
\begin{tabular}{ll|ll||ll}    \toprule $\mathrm{Alice}_1$ & Bob  & $\mathrm{Alice}_2$ & Bob & Charlie & Bob  \\\toprule
$\ket{\Phi^+}$ & & $\ket{\Phi^+}$ & & $\ket{0}\ket{+}$ & \\
$\ket{\Psi^+}$ & $X_3$ & $\ket{\Psi^+}$ & $X_4$ & $\ket{0}\ket{-}$ & $Z_3Z_4$\\
$\ket{\Phi^-}$ & $Z_3$ & $\ket{\Phi^-}$ & $Z_4$ & $\ket{1}\ket{+}$ & $X_3X_4$\\
$\ket{\Psi^-}$ & $Z_3X_3$ & $\ket{\Psi^-}$ & $Z_4X_4$ & $\ket{1}\ket{-}$ & $Z_3Z_4X_3X_4$\\
\bottomrule 
 \hline 
\end{tabular}}
\caption{Correction operators for a two-qubit controlled quantum teleportation scheme using $\ket{\mathrm{T}}$ as the entangled resource state. }
\label{correction-ops}
\end{table}
If Charlie wished to complete the teleportation, he would measure one of his qubits in the $Z$ basis and the other in the $X$ basis. He would then send the measurement outcomes to Bob, who would conditionally apply further corrections to his qubits according to Table \ref{correction-ops}. When Bob is given both Alice's and Charlie's measurement outcomes, the fidelity of the teleportation protocol is unity. With only Alice's information, however, Bob's qubits are in a mixed state $\rho_{\mathrm{B}}$ obtained by tracing over Charlie's qubits. The fidelity $\langle\varphi\lvert \rho_{\mathrm{B}}\rvert \varphi\rangle$ of the teleportation protocol for state $\ket{\varphi}$ can then be averaged over the Haar measure $d\varphi$ to obtain the average fidelity
\begin{equation}
    \bar{F}=\int d\varphi \:\langle\varphi\lvert \rho_{\mathrm{B}}\rvert \varphi\rangle=\frac{2}{5},
\end{equation}
corresponding to a control power of $1-\bar{F}=3/5$. The average teleportation fidelity that Alice and Bob achieve without Charlie's help is therefore only equal to the fidelity of the best classical strategy~\cite{popescu1994bell}, in which Alice performs an optimal measurement of her two-qubit state and sends the result to Bob, who then attempts to reconstruct Alice's state on the basis of this classical information. The maximal fidelity for this classical strategy is given by the optimal fidelity for estimation of a quantum state, equal to $2/5$ for a two-qubit system~\cite{bruss1999optimal, acin2000optimal,badziag2000local}. Charlie therefore has the power to prevent Alice and Bob from deriving any advantage (over classical strategies) from their shared entanglement.

We note that the notion of control power given above only has meaning when the three parties share a pure state. Noise can, however, be incorporated under appropriate conditions \cite{gangopadhyay2022controlled}. Whether the noisy tetrahedron state resulting from a sequence of flying-cat parity checks with finite measurement errors and finite photon loss provides sufficient control power is a question for future study, beyond the scope of the present work.

\section{Feasibility}\label{feasibility}

In this section, we consider the feasibility of implementing flying-cat parity checks in a state-of-the-art circuit QED architecture \cite{blais2021circuit}, using currently realized experimental parameters. The basic elements of the parity check consist of: (i) an entangling operation realizing the qubit-state-dependent phase shift of Eq.~\eqref{entangle}, and (ii) propagation of the microwave-frequency light pulse through superconducting transmission lines and circulators [Fig.~\ref{fig:tetrahedron}(b)]. We begin by discussing error sources related to (i), namely internal cavity losses and errors due to a finite pulse duration, and then errors due to qubit decoherence.

\subsection{Entangling operation}

To assess the quality of the entangling operation, here we focus on a circuit-QED setup consisting of a single-sided microwave cavity with a mode at frequency $\omega_\mathrm{c}$, coupled to a transmission line with coupling rate $\kappa_0$. In addition, the cavity is coupled via a transverse (Rabi) coupling $g$ to a two-level system encoding a qubit with level splitting $\omega_\mathrm{q}$. As in Fig.~\ref{fig:phase-shift}, we consider a regime where an incident microwave pulse prepared in a coherent state $\ket{\alpha}$ acquires a qubit-state-dependent phase shift upon reflection [$\alpha\to (-1)^s\alpha$ for $s=0,1$]. However, in contrast to the setup in Fig.~\ref{fig:phase-shift}, which relies on strong, resonant coupling and the existence of a third level, here we focus on an alternative approach due to its compatibility with qubits having no nearby third level, e.g., electron-spin qubits~\cite{mi2018coherent,samkharadze2018strong}. A similar analysis could also be carried out for the setup in Fig.~\ref{fig:phase-shift}, leading to slightly different conditions.

In the dispersive regime of cavity QED ($|\delta|\gg |g|$, with detuning $\delta=\omega_\mathrm{q}-\omega_\mathrm{c}$), the cavity and qubit are described approximately by the effective Hamiltonian \cite{blais2004cavity} (setting $\hbar=1$ here and throughout):
\begin{equation}
    H=\omega_\mathrm{c} a^\dagger a+\frac{1}{2}\left(\omega_\mathrm{q}+\chi\right)\sigma_z+\chi a^\dagger a\sigma_z.
\end{equation}
Here, $\chi=g^2/\delta$ is the dispersive shift, $a$ annihilates a photon in the cavity, and $\sigma_z=\ketbra{0}{0}-\ketbra{1}{1}$. In the absence of bit-flips, the Pauli operator $\sigma_z$ is preserved and can be replaced by $\sigma_z\mapsto (-1)^s$ for qubit state $s$. Under the standard (Born-Markov) approximation for a wide-bandwidth transmission line, the cavity field evolves according to the quantum Langevin equation
\begin{equation}\label{eq:Langevin}
\dot{a}(t)=i\comm{H}{a(t)}-\frac{1}{2}\left(\kappa_0+\kappa_\mathrm{int}\right)a(t)-\sqrt{\kappa_0}r_\mathrm{in}(t),
\end{equation}
where $r_\mathrm{in}(t)$ characterizes the complex amplitude of the incoming microwave pulse, and where $\kappa_\mathrm{int}$ gives the rate of internal cavity losses. From Eq.~\eqref{eq:Langevin} and the input-output relation~\cite{gardiner1985input}, $r_\mathrm{out}(t)=(2\pi)^{-1}\int d\omega e^{-i\omega t} r_\mathrm{out}(\omega)=r_\mathrm{in}(t)+\sqrt{\kappa_0}\left<a(t)\right>$, we then find the qubit-state dependent reflection coefficient $R_s(\omega)=r_\mathrm{out}(\omega)/r_\mathrm{in}(\omega)$:
\begin{equation}\label{reflection-coefficient}
R_s(\omega) =\frac{2i[\omega-\omega_\mathrm{c}-(-1)^s\chi]+\kappa_0-\kappa_\mathrm{int}}{2i[\omega-\omega_\mathrm{c}-(-1)^s\chi]-\kappa_0-\kappa_\mathrm{int}}.
\end{equation}
For an ideal narrow-bandwidth input pulse with $\omega\simeq \omega_\mathrm{c}$, and for negligible internal losses, $\kappa_\mathrm{int}\simeq 0$, a special choice of dispersive shift $\lvert\chi\rvert=\kappa_0/2$ leads to a phase shift $\alpha\mapsto R_s(\omega)\alpha\simeq (-1)^s i\alpha$ \cite{kono2018quantum,wang2022flying}. The operation of Eq.~\eqref{entangle} can then be recovered with a change in phase reference, $\alpha\mapsto -i\alpha$. Deviations from the idealized assumptions above (e.g., finite-bandwidth pulses and finite internal losses $\kappa_\mathrm{int}\ne 0$) will generally lead to an imperfect entangling operation.

As a measure of quality for the entangling operation, we use the infidelity $\varepsilon$ of the final joint state $\rho_{\mathrm{imperfect}}$ of the qubit and microwave pulse obtained in the presence of imperfections, relative to the ideal final state $\ket{\psi_{\mathrm{ideal}}}=(\ket{0,\alpha}+\ket{1,-\alpha})/\sqrt{2}$ for a qubit initially prepared in $\ket{+}$:
\begin{equation}
  \varepsilon=1-\langle \psi_{\mathrm{ideal}}\vert \rho_{\mathrm{imperfect}}\vert \psi_{\mathrm{ideal}}\rangle\approx \varepsilon_{\mathrm{qubit}}+\varepsilon_{\mathrm{reflect}}.  
\end{equation}
Here, we have divided $\varepsilon$ into a contribution $\varepsilon_{\mathrm{qubit}}$ due to qubit decoherence (neglecting other error sources) and a contribution $\varepsilon_{\mathrm{reflect}}$ due to the combined effects of internal cavity losses and finite pulse bandwidth (neglecting decoherence).

To quantify the effects of both finite $\kappa_{\mathrm{int}}$ and finite bandwidth $\sim \tau^{-1}$ (for a pulse with duration $\tau$), we assume that the spatiotemporal mode supporting state $\ket{\alpha}$ has waveform $u(t)$, normalized according to $\int dt\lvert u(t)\rvert^2=1$~\cite{wang2005engineering}. Before the reflection, frequency mode $\omega$ of the transmission line is in a coherent state having amplitude $\alpha u(\omega)$, where here, $u(\omega)$ is the Fourier transform of $u(t)$. For an ideal entangling operation, the waveform would acquire a well-defined $\pm \pi/2$ phase shift, $u(\omega)\mapsto \pm i u(\omega)$, conditioned on the qubit state, in which case we could simply write $\ket{\alpha}\mapsto\ket{\pm i \alpha}$. However, in practice, $\alpha u(\omega)\mapsto \alpha R_s(\omega)u(\omega)$. Evaluating the overlap of the ideal output state with the state resulting from a finite $\kappa_{\mathrm{int}}$ and finite $\tau$ gives 
\begin{align}
\begin{aligned}\label{eq:epsilon-reflect}
\varepsilon_{\mathrm{reflect}}= 1-\frac{1}{2}\sum_{s=0,1}e^{{-}\alpha^2\int\frac{d\omega}{2\pi}\lvert u(\omega)\rvert^2\lvert (-1)^si-R_s(\omega)\rvert^2}.
\end{aligned}
\end{align}
For a Gaussian waveform with width $\tau$, corresponding in the frequency domain to $\lvert u(\omega)\rvert^2=2\sqrt{\pi}\tau e^{-(\omega-\omega_{\mathrm{c}})^2\tau^2}$, the integrand in Eq.~\eqref{eq:epsilon-reflect} has finite weight only for $\lvert \omega-\omega_{\mathrm{c}}\rvert\lesssim 1/\tau$. Expanding $R_s(\omega)$ to leading nontrivial order in both $\kappa_{\mathrm{int}}/\lvert\chi\rvert$ and $|\omega-\omega_\mathrm{c}|/\lvert\chi\rvert\lesssim 1/(\tau\lvert\chi\rvert)$ then gives
\begin{equation}
\left|(-1)^s i-R_s(\omega)\right|^2\simeq \frac{1}{4}\left(\frac{\kappa_\mathrm{int}}{\chi}\right)^2+\left(\frac{\omega-\omega_{\mathrm{c}}}{\chi}\right)^2.
\end{equation}
Inserting this result into Eq.~\eqref{eq:epsilon-reflect}, performing the Gaussian integral, then expanding to leading nontrivial order in $\kappa_\mathrm{int}/\lvert\chi\rvert$ and $1/(\tau\lvert\chi\rvert)$ gives
\begin{equation}\label{eq:epsilon-reflect-2}
    \varepsilon_{\mathrm{reflect}}\simeq \alpha^2\left(\frac{1}{2\tau^2\lvert\chi\rvert^2}+\frac{\kappa_{\mathrm{int}}^2}{4\lvert\chi\rvert^2}\right).
\end{equation}
Taking a large dispersive shift therefore has two benefits: first, reducing the effects of internal cavity loss $\sim \kappa_{\mathrm{int}}$, and second, allowing for a faster error-correction cycle $\sim\tau$ while maintaining a small error, $\varepsilon_{\mathrm{reflect}}$. 

The setup described above has been realized experimentally in, e.g., Ref.~\cite{wang2022flying} for a transmon qubit coupled to a microwave cavity. For the values realized in Ref.~\cite{wang2022flying} ($\kappa_{\mathrm{int}}/2\pi=0.22$ MHz, $\chi/2\pi=-1.05$ MHz, and $\tau=500$ ns), and assuming $\alpha=1$, Eq.~\eqref{eq:epsilon-reflect-2} gives a contribution to the infidelity $\simeq 0.046$ from the term $\propto 1/\tau^2$ and a contribution $\simeq 0.004$ arising from internal losses. The estimate above therefore suggests $\varepsilon_{\mathrm{reflect}}\approx 0.05$, dominated by the finite-bandwidth pulses, although it should be noted that the pulse used in Ref.~\cite{wang2022flying} was square, rather than Gaussian. Doubling the value of the dispersive shift (so that $\chi/2\pi=-2.10$ MHz) would reduce the error $\propto \tau^{-2}$ to 0.01 for the same value of $\tau=500$ ns, comparable to the $O(1\%)$ threshold of the subsystem surface code~\cite{bravyi2012subsystem,higgott2021subsystem}.  A much smaller value of $\kappa_{\mathrm{int}}/2\pi\sim 100\,\mathrm{Hz}$ also remains well within the state of the art~\cite{reagor2016quantum}. Therefore, we do not expect $\varepsilon_{\mathrm{reflect}}$ to be a limiting error source for near-term implementations.

The expression for the reflection coefficient $R_s(\omega)$ [Eq.~\eqref{reflection-coefficient}] neglects the effects of qubit decoherence, which will also degrade the quality of the entangling operation.  To guarantee a low probability of qubit decoherence on the timescale $\tau$ of the pulse, we require that $\mathrm{min}\{T_1,T_2^*\}\gg \tau$, where here, $T_1$ is the timescale for energy relaxation and $T_2^*$ is the inhomogeneous dephasing time due to random fluctuations in the qubit splitting from shot to shot. For dephasing caused by a Markovian (short-correlation-time) environment, the homogeneous dephasing time $T_2$ is equal to $T_2^*$, and the decay of coherences is exponential with timescale $T_2=T_2^*$. In Ref.~\cite{wang2022flying}, decoherence was limited by pure dephasing on a time scale $T_2=T_2^*=6$ $\mu$s, while the pulse had a duration of $\tau=500$ ns. To leading order in $\tau/T_2^*$, a quick estimate for the infidelity due to decoherence is then given by 
\begin{equation}
    \varepsilon_{\mathrm{qubit}}\simeq \tau/T_2^*.
\end{equation}
Inserting the values for $\tau$ and $T_2^*$ from Ref.~\cite{wang2022flying} then gives $\varepsilon_{\mathrm{qubit}}\simeq0.08$, comparable to the infidelity of 0.11 attributed to the effects of decoherence, based on a more sophisticated error model \cite{wang2022flying}. Reducing the pulse duration therefore provides a clear avenue towards reducing errors due to decoherence. However, to maintain the same small value of $\varepsilon_{\mathrm{reflect}}$, a reduction in $\tau$ must be accompanied by a proportional increase in $\vert \chi\rvert$, as discussed above.

\subsection{Transmission losses}

For the setup illustrated in Fig.~\ref{fig:tetrahedron}(b), sources of photon loss will likely include losses $\eta_{\mathrm{trans}}$ incurred in transit, together with losses $\eta_{\mathrm{circ}}$ due to chiral elements like circulators, giving
\begin{equation}
    \label{losses}\eta=\eta_{\mathrm{trans}}+\eta_{\mathrm{circ}}.
\end{equation}
Many superconducting interconnects are made of NbTi, for which typical attenuation rates are 5 dB/km, corresponding to losses of 68\%/km~\cite{niu2023low}. The architecture of Ref.~\cite{niu2023low} instead uses pure Al to achieve losses of 0.15 dB/km (3.4\%/km), comparable to the values reached for fiber-optic cables  carrying telecom-frequency light. Hence, for a meter-scale experiment, $\eta_{\mathrm{trans}}$ can likely be made negligible relative to other sources of loss. 

In microwave platforms, chiral elements like circulators often lead to higher losses than those incurred during free propagation, due in part to the challenges of on-chip integration. In Ref.~\cite{storz2023loophole}, for instance, circulators led to insertion losses estimated at $\eta_{\mathrm{circ}}=0.13$. If the dominant source of error is circulator loss, a factor-of-two improvement would be sufficient to demonstrate multipartite entanglement in a tetrahedron state using the layout of Fig.~\ref{fig:tetrahedron}(b), since $\langle\mathcal{W}\rangle<0$ can be achieved for the entanglement witness $\mathcal{W}$ [Eq.~\eqref{witness}] provided $\eta \lesssim 0.05$ [cf.~Fig.~\ref{fig:witness}]. Significant effort has gone towards the development of microwave circulators compatible with on-chip integration~\cite{zhang2021charge,navarathna2023passive}, as this technology has already been identified as a key ingredient for scaling up superconducting processors. As a final comment, the circulators shown in Fig.~\ref{fig:tetrahedron}(b) are not strictly necessary. They could be replaced by fast dynamical switches that  control the path taken by the pulse.

\subsection{Summary: Parametric constraints}

For the circuit QED architecture considered above, where the entangling operation of Eq.~\eqref{entangle} is realized in the dispersive regime, increasing the strength of the dispersive coupling provides a straightforward avenue towards increasing the fidelity of the entangling operation. Decoherence times are another important consideration. For $\mathrm{min}\{T_1,T_2^*\}=10\:\mu s$, and for the same value of $\kappa_{\mathrm{int}}/2\pi=0.22$ MHz realized in Ref.~\cite{wang2022flying}, a possible combination of $\tau$ and $\chi$ giving $\varepsilon\approx\varepsilon_{\mathrm{qubit}}+\varepsilon_{\mathrm{reflect}}\approx 0.01$ is $\tau=100$ ns and $\lvert \chi\rvert/2\pi=10$ MHz. Dispersive shifts as large as $10$ MHz have been realized experimentally for transmon qubits~\cite{noh2023strong}.

Assuming the layout of Fig.~\ref{fig:tetrahedron}(b), preparing a tetrahedron state requires six stabilizer measurements, each involving three entangling operations, for a total of eighteen entangling operations. Accounting for errors in the entangling operations, an estimate for the maximum achievable preparation fidelity of the tetrahedron state is therefore $F\le F_\mathrm{max}\simeq 1-18\varepsilon$. Since a fidelity $F>1/2$ is required to demonstrate multipartite entanglement [Fig.~\ref{fig:witness}], an ideal scenario is one where all error sources associated with the entangling operation (e.g.~qubit dephasing, internal cavity losses, and finite-bandwidth effects) can be controlled to the extent that $18\varepsilon \ll 1/2$ ($\varepsilon \ll 0.03$). If this condition is met, then multipartite entanglement can be demonstrated with a photon loss rate as high as $\eta\approx 0.05$ [Fig.~\ref{fig:witness}]. As discussed above, this would require only a factor $\sim 2$ improvement in the circulator insertion losses reported in Ref.~\cite{storz2023loophole}. 

\section{Conclusion}

High-quality flying-cat parity checks could provide the ability to perform error correction in distributed architectures, where native qubit-qubit interactions are not typically available across nodes. However, operations involving propagating quasimodes of the electromagnetic field will introduce photon loss as a source of error on code qubits. 

In this work, we have analyzed the trade-off between measurement errors and errors due to photon loss in the context of weight-3 flying-cat parity checks. We have shown how the total impact of these error sources could be minimized as a function of the coherent-state amplitude. In addition, we have verified that the horizontal hook errors due to photon loss remain gauge-equivalent to single-qubit errors for gauge operators of the form $X^{\otimes 3}$, $Z^{\otimes 3}$, so that flying-cat parity checks preserve this key advantage of the subsystem surface code. In the context of entanglement distribution, we also described how flying-cat parity checks can be used to generate three-qubit GHZ states, as well as a six-qubit entangled ``tetrahedron'' state that can be used for controlled quantum teleportation of an arbitrary two-qubit state. Another result of this paper is the conditional post-measurement state $\rho_x$ of the data qubits following a phase-sensitive measurement of the electromagnetic field, incorporating both measurement errors and photon-loss-induced errors. This object can be used to find the conditional probabilities for error patterns on the code qubits, shot-by-shot for each measured value of $x$. In a quantum error correcting code, this probability could be passed to a decoding scheme that takes full advantage of this soft information \cite{fukui2017analog,fukui2018high,pattison2021improved}. Finally, we have shown that high-quality flying-cat parity checks are feasible with current or near-future technology. 

Flying-cat parity checks rely on sources and detectors for (classical) coherent states of light, rather than the single photons used to link nodes in several other quantum-network implementations. These parity checks therefore do not require single-photon sources or detectors, although they do require a high-quality stable source of coherent light and low-noise phase-sensitive measurements. We have only considered the simplest encoding (a single-mode input coherent state). In future work, it may be interesting to consider whether a more elaborate encoding could be used to simultaneously detect photon loss and parity using, e.g., a cat-code variant \cite{bergmann2016quantum}. Additionally, rapid qubit gates and dynamical decoupling sequences performed to preserve code qubits coupled to cavities will generically lead to bursts of light that can carry away information~\cite{mcintyre2022non}. Whether dynamical decoupling can be made compatible with flying-cat parity checks is an open question. 

The flexible qubit connectivity enabled by the use of propagating light pulses could also provide an avenue towards realizing the nonlocal interactions required by certain quantum low-density parity-check (LDPC) codes~\cite{bravyi2010tradeoffs,breuckmann2021quantum,cohen2022low,baspin2022quantifying, bravyi2023high}. Quantum LDPC codes have shown promise for reducing the number of physical qubits required to encode a logical qubit compared to leading candidates like the surface code. 

\begin{acknowledgments}
We thank K.~Wang for useful discussions. We also acknowledge funding from the Natural Sciences and Engineering Research Council of Canada (NSERC) and from the Fonds de Recherche du Qu\'ebec--Nature et technologies (FRQNT). 
\end{acknowledgments}

\end{document}